\documentstyle[aps,12pt,manuscript]{revtex}
\tightenlines

\begin{document}
\newcommand{\beq}{\begin{equation}}
\newcommand{\eeq}{\end{equation}}
\newcommand{\beqa}{\begin{eqnarray}}
\newcommand{\eeqa}{\end{eqnarray}}
\newcommand{\fr}{\frac}
\draft
\preprint{INJE-TP-02-02, hep-th/0204083}

\title{Dynamical Behavior of dilaton in  de Sitter space}

\author{ H. W. Lee and Y. S. Myung}
\address{Relativity Research Center and School of Computer Aided Science,
Inje University, Gimhae 621-749, Korea}

\maketitle

\begin{abstract}
We study the dynamical behavior of the dilaton in the background of three-dimensional
Kerr-de Sitter space
which is  inspired from  the
low-energy string effective action.
The perturbation analysis around the cosmological horizon of Kerr-de
Sitter space
reveals a mixing between the dilaton and other fields.  Introducing
a gauge (dilaton gauge), we  can disentangle this mixing completely and obtain
one decoupled dilaton equation. However it turns out that this belongs to the
tachyon. The stability of  de Sitter solution with $J=0$ is discussed.
Finally we compute the dilaton  absorption cross section to extract
information on the cosmological horizon of de Sitter space.
\end{abstract}

\newpage
\section {introduction}
Recently an accelerating universe has proposed to be a way
to interpret the astronomical data of supernova\cite{Per,CDS,Gar}.
The inflation is employed to solve the cosmological flatness and
horizon puzzles arisen in the standard cosmology.
Combining this observation with the need of inflation
 leads to that our universe approaches de Sitter
geometries in both the infinite past and the infinite future\cite{Wit,HKS,FKMP}. Hence it is
very important to study the nature of de Sitter (dS) space and the
dS/CFT correspondence\cite{BOU}.
 However,
there exist  difficulties in studying de Sitter space.
First there is no spatial infinity and   global timelike
Killing vector.  Thus it is not easy to define the conserved  quantities including  mass,
charge and angular momentum appeared in asymptotically  de Sitter space.
Second the dS solution is absent from string theories and thus we
do not   have a definite example  to test the dS/CFT correspondence.
Finally it is hard to define  the $S$-matrix because of the
presence of the cosmological horizon.

Accordingly most of works on  de Sitter space were concentrated on
the massive scalar propagation and its quantization\cite{STR,BMS,SV,YS}. Also the
bulk-boundary relation for the scalar was introduced to study the
dS/CFT correspondence\cite{AWLQ}. Hence it is important to find a model which
can accommodate  the de Sitter space solution. In this work we
introduce an interesting model which is motivated from the
low-energy string action in (2+1)-dimensions\cite{Hor}. This model includes a
nontrivial scalar so-called the dilaton\footnote{Previously we are interested
in anti de Sitter black hole like the BTZ black hole\cite{BTZ}.
In the BTZ black hole and the three-dimensional black string,
the role of the dilaton was discussed in ref.\cite{flux}.}
. Actually we will use the
dilaton to investigate the nature of the cosmological horizon in
de Sitter space.

It is known that the cosmological horizon is very similar
to the event horizon in the sense that one can define its
thermodynamic quantities using the same way as is done for
the black hole. Two important quantities to understand the black hole
are the Bekenstein-Hawking entropy and the absorption cross
section (greybody factor). The former specifies   intrinsic
property of the black hole itself, while the latter relates to the
effect of spacetime curvature surrounding the black hole.
We emphasized here the greybody factor for
the black hole arises as a consequence of scattering off the
gravitational potential surrounding the event horizon\cite{grey1}. For example,
the low-energy $s$-wave greybody factor for a massless scalar has a
universality such that it
is equal to the area of the horizon for all spherically symmetric
 black holes\cite{grey2}. This can be obtained  by solving the wave
 equation explicitly.  The entropy for the cosmological horizon was
 first discussed in literature\cite{entropy}. However,
 there exist
 a few of the wave equation approaches to find the greybody factor  for the
 cosmological horizon\cite{YS}.
 A similar work for the four-dimensional
 Schwarzschild-de Sitter black hole appeared  in\cite{KOY} but it
 focused mainly on obtaining the temperature of  the eternal  black hole.
Also the absorption rate for the four-dimensional Kerr-de Sitter
black hole was discussed in\cite{STU}.

In this paper we compute the absorption cross section of the
dilaton in the background of  three-dimensional de Sitter (dS$_3$) space
with the cosmological horizon.
For this purpose we first confine  the wave equation only to the southern
diamond where the time evolution of  waves is properly defined even if this area
does not include spatial infinity.
And then we
solve the wave equation to find the greybody factor in the low-energy and
low-temperature limits.

The organization of this paper is as follows. In section II we
briefly review our model inspired from the low-energy string action and
its  Kerr-de Sitter solution. We
introduce
the perturbation to study   the cosmological horizon in the
background of Kerr-de Sitter space in section III.
In section IV we perform the potential analysis to check whether
the de Sitter solution is or not stable.
In section V we calculate  the absorption cross section
 for $j=1,2$-angular modes of the dilaton explicitly.
Finally we discuss our results in section VI.

\section{Kerr-de Sitter solution}

We start with the low-energy string action in string frame\cite{Hor}
\begin{equation}
S_{l-s} = \int d^3 x \sqrt{-g} e^{\Phi}
   \big \{ R + (\nabla \Phi)^2 + {8 \over k} - {1 \over 12} H^2  \big \},
\label{action}
\end{equation}
where $\Phi$ is the dilaton, $H_{\mu\nu\rho}=3 \partial_{[\mu}B_{\nu\rho]}$
is the Kalb-Ramond field, and $k$ the cosmological constant.
This action was widely used for studying the BTZ black hole and
the black string\cite{flux}.
Although $k$ was originally proposed to be positive, here we assume to extend it
to be negative for our purpose.
The equations of motion lead to
\begin{eqnarray}
R_{\mu\nu} - \nabla_\mu \nabla_\nu \Phi
-{1\over 4} H_{\mu \rho \sigma} H_\nu^{\rho \sigma} &=& 0,
\label{eq_graviton} \\
\nabla^2 \Phi + (\nabla \Phi)^2 - {8 \over k} - {1 \over 6} H^2   &=& 0,
\label{eq_scalar1}  \\
 \nabla_\mu H^{\mu \nu \rho} + (\nabla_\mu \Phi) H^{\mu \nu \rho} &=& 0.
\label{eq_anti}
\end{eqnarray}
The Kerr-de Sitter  solution to
Eqs.(\ref{eq_graviton})-(\ref{eq_anti}) is found to be\cite{BMS}
\begin{eqnarray}
&& \bar H_{txr} =-i\fr{2r}{\ell}, ~~~~~~\bar
\Phi = 0,~~~~~~k=-2 \ell^2,
      \nonumber   \\
&& \bar g_{\mu\nu} =
 \left(  \begin{array}{ccc}  -(M - {r^2 / \ell^2}) & -{J / 2} & 0  \\
                             -{J / 2} & r^2 & 0  \\
    0 & 0 & f^{-2}
         \end{array}
 \right)
\label{bck_metric}
\end{eqnarray}
with the metric function $f^2 =M -r^2 / \ell^2  +J^2 / 4 r^2$.
The above Kerr-de Sitter solution is obtained from the BTZ black hole
\cite{BTZ} by replacing both $M$ and $\ell^2$
by $-M$ and $-\ell^2$.
The metric $\bar g_{\mu\nu}$ is singular at $r=r_{\pm}$,
\begin{equation}
r_{\pm}^2 = {{M\ell^2} \over 2} \left \{ 1 \pm \left [
   1 + \left ( {J \over M\ell} \right )^2 \right ]^{1/2} \right \}
\label{horizon}
\end{equation}
with $M=(r_+^2 + r_-^2) / \ell^2=(r_+^2-r_{(-)}^2)$ and $ J=2 r_+r_{(-)} / \ell$.
For convenience we introduce $r_{(-)}^2\equiv -r^2_->0$ due to $r^2_-<0$
in Kerr-de Sitter space.
We note that in (2+1)-dimensions, there is no black hole horizon
for Kerr-de Sitter space because the black hole degenerates to a
conical singularity at the origin $r=0$. This singularity gives rise to
some difficulties to analyze the wave equation in the southern diamond of Kerr-de Sitter
space. In this work we consider the
cosmological horizon $r_c=r_+$ with interest.
For convenience, we list the Hawking temperature $T_c$, the area of
the cosmological horizon ${\cal A}_c$, and the angular velocity at the horizon
$\Omega_c$ as
\begin{equation}
T_c = (r_+^2 + r_{(-)}^2) / 2 \pi \ell^2 r_+,
~~{\cal A}_c = 2 \pi r_+,
~~\Omega_c = J / 2 r_+^2.
\label{temp}
\end{equation}
For $J=0$ case, one
finds de Sitter solution which gives us with $r_+=\ell$ and $r_-=0$
\begin{equation}
M=1,~~J=0,~~T_c = 1 / 2 \pi \ell,
~~{\cal A}_c = 2 \pi \ell,
~~\Omega_c =0.
\label{tempd}
\end{equation}

\section{ Perturbation around Kerr-de Sitter solution}

To study the propagation of all fields in  Kerr-de Sitter space specifically,
 we introduce
the small perturbation fields\cite{flux}
\begin{equation}
H_{t \phi r} = \bar H_{t \phi r} + {\cal H}_{t \phi r},~~~
\Phi = 0 + \varphi,~~~
g_{\mu\nu} = \bar g_{\mu\nu} + h_{\mu\nu}
\label{per}
\end{equation}
around the background solution of
Eq.(\ref{bck_metric}).
For convenience,
we introduce the notation
${\hat h}_{\mu\nu} = h_{\mu\nu}- {\bar g}_{\mu\nu} h/2$ with
$h= h^\rho_{~\rho}$.
And then one needs to linearize Eqs.(\ref{eq_graviton})-(\ref{eq_anti})
to obtain
\begin{eqnarray}
  \delta R_{\mu\nu} (h)
- \bar \nabla_\mu \bar \nabla_\nu \varphi
- {1 \over 2} \bar H_{\mu \rho \sigma} {\cal H}_\nu^{~ \rho \sigma}
+ {1 \over 2} \bar H_{\mu \rho \sigma} \bar H_{\nu\alpha}^{~~\sigma}
h^{\rho \alpha} &=& 0,
\label{lin_graviton} \\
 \bar \nabla^2 \varphi
- {1 \over 6} \Big \{ 2 \bar H_{\mu \rho \sigma}
                      {\cal H}^{\mu \rho \sigma}
     - 3 \bar H_{\mu \rho \sigma}   \bar H^{\alpha \rho \sigma} h^\mu_\alpha
                                    \Big \} &=& 0,
\label{lin_scalar} \\
   \bar \nabla_\mu  {\cal H}^{\mu \nu \rho}
- ( \bar \nabla_\mu h_\beta^{~\nu}) {\bar H}^{\mu\beta\rho}
+ (\bar \nabla_\mu h_\beta^{~\rho}) {\bar H}^{\mu\beta\nu}
- (\bar \nabla_\mu {\hat h}_{~\alpha}^{\mu}) {\bar H}^{\alpha\nu\rho}
      + (\partial_\mu \varphi) \bar H^{\mu \nu \rho}
 &=& 0,
\label{lin_anti}
\end{eqnarray}
where the Lichnerowicz operator $\delta R_{\mu\nu}(h)$
is given by
\begin{eqnarray}
&&\delta R_{\mu\nu} = -{1 \over 2} \bar \nabla^2 h_{\mu\nu}
 +{\bar R}_{\sigma ( \nu} h^\sigma_{~\mu )}
 -{\bar R}_{\sigma \mu\rho\nu} h^{\sigma\rho}
 + \bar \nabla_{( \nu} \bar \nabla_{|\rho|} {\hat h}^\rho_{~\mu)}.
\label{delR}
\end{eqnarray}
These are the bare perturbation equations.  It is desirable to examine whether
we  make a choice of perturbation and gauge which can simplify
Eqs.(\ref{lin_graviton})-(\ref{lin_anti}) significantly.
For this purpose we wish to count the physical degrees of freedom.
A symmetric
traceless tensor has D(D+1)/2--1 in D-dimensions.  D of them
are eliminated by the gauge condition.  Also D--1 are
eliminated from our freedom to take
further residual gauge transformations.
Thus gravitational degrees of freedom are D(D+1)/2--1--D--(D--1)=D(D--3)/2.
In three dimensions we have no
propagating degrees of freedom for $h_{\mu\nu}$.
Also two-form $B_{\mu\nu}$ has no physical degrees of freedom for D=3.
Hence the physical degree of freedom in the Kerr-de Sitter solution
is just the dilaton $\varphi$.

Considering the $t$ and  $\phi$-symmetries of the background
spacetime Eq.(\ref{bck_metric}),
we can decompose $h_{\mu\nu}$ into frequency ($\omega$) and angular ($j=0,1,2,\cdots$)
modes in these
variables
\begin{equation}
h_{\mu\nu}(t,\phi,r) = e^{-i \omega t} e^{ij \phi}H_{\mu\nu}(r).
\label{ptr_metric}
\end{equation}
For simplicity, one chooses the same perturbation as in Eq.(\ref{ptr_metric})
 for Kalb-Ramond field and dilaton as
\begin{eqnarray}
{\cal H}_{t \phi r}(t,\phi,r) &&= \bar H_{t \phi r} {\cal H}(t,\phi,r)
 =\bar H_{t \phi r} e^{-i \omega t} e^{i j \phi} \tilde{\cal H}(r),
\label{ptr_anti} \\
\varphi(t,\phi,r)&&=e^{-i \omega t} e^{i j \phi} \tilde\varphi(r).
\label{ptr_scalar}
\end{eqnarray}
Since the dilaton is only a propagating mode, we are
interested in the dilaton equation (\ref{lin_scalar}).
Note that Eq.(\ref{lin_graviton}) is irrelevant to our analysis, because
it belongs to the redundant relation.
Eq.(\ref{lin_scalar}) can be rewritten as
\begin{eqnarray}
\bar \nabla^2 \varphi
+{4 \over l^2} (h - 2 {\cal H} )
=0.
\label{eq_scalar}
\end{eqnarray}
If we start with full six degrees of freedom of Eq.(\ref{ptr_metric}), we
should choose a gauge.
Conventionally, we choose the
harmonic  gauge ($\bar \nabla_\mu {\hat h}^{\mu\rho} = 0$) to
describe the propagation of gravitons in D$>$3 dimensions\cite{wein}.
It turns out that a mixing between the dilaton and other fields of $h,{\cal H}$ is
not disentangled completely with the
harmonic gauge condition. Here we focus on the propagation of the dilaton $\varphi$.
Fortunately if we introduce the
dilaton gauge
($\bar\nabla_\mu \hat h^{\mu \rho}=h^{\mu \nu} \Gamma^\rho_{\mu \nu} $),
this difficulty may be resolved.
Actually this gauge was designed for the dilaton propagation\cite{dilga}.
We attempt to disentangle the last term in Eq.(\ref{eq_scalar}) by
using both the dilaton gauge and
Kalb-Ramond equation (\ref{lin_anti}).
Each component ($\rho=t,\phi,r$) of the  dilaton gauge condition gives rise to
\begin{eqnarray}
t&:& (\partial_r + {1 \over r} ) h^{tr}
- i \omega h^{tt} + i j h^{t \phi}
+ {1 \over 2} i \omega h \bar g^{tt} - {1 \over 2}i j h \bar g^{t\phi} = 0,
\label{eq_gauge_t}  \\
\phi&:& (\partial_r + {1 \over r} ) h^{\phi r}
- i \omega h^{\phi t} + i j h^{\phi \phi}
+ {1 \over 2}i \omega h \bar g^{\phi t} - {1 \over 2}i j h \bar g^{\phi \phi} = 0,
\label{eq_gauge_x}  \\
r&:& (\partial_r + {1 \over r} ) h^{rr}
- i \omega h^{rt} + i j h^{r \phi}
- {1 \over 2} (\partial_r h) \bar g^{rr} = 0.
\label{eq_gauge_r}
\end{eqnarray}
And each component ($\nu,\rho$) of the Kalb-Ramond equation (\ref{lin_anti}) leads to
\begin{eqnarray}
t\phi:&& -\partial_r (\varphi+{\cal H} -{h^t_{~t}} -{h^\phi_{~\phi}})
+ {1 \over rf^2}\left (-M +{3 r^2 \over \ell^2} +
                     {J^2 \over 4r^2}\right ) h^r_{~r}
+i\omega h^t_{~r} -i j h^\phi_{~r}=0,
\label{eq_anti_tx} \\
tr:&& -i j (\varphi+{\cal H} -h^t_{~t} -h^r_{~r})
- {1 \over r} h^r_{~\phi}
+2 f^2 h^\phi_{~r}
-\partial_r h^r_{~\phi} +i\omega h^t_{~\phi} =0,
\label{eq_anti_tr} \\
\phi r:&& -i\omega (\varphi+{\cal H} -h^\phi_{~\phi} -h^r_{~r})
           + {1 \over r} h^r_{~t}
-{2 rf^2 \over \ell^2} h^t_{~r}
+\partial_rh^r_{~t} +i j h^\phi_{~t} =0.
\label{eq_anti_xr}
\end{eqnarray}
Solving six equations (\ref{eq_gauge_t})-(\ref{eq_anti_xr}), one finds an
important constraint
\begin{equation}
\partial_\mu (2 \phi +2 {\cal H} - h ) =0, ~~\mu=t,\phi,r
\label{eq_anti_simple}
\end{equation}
which leads to
$h - 2 {\cal H} =2 \varphi$.
This means that $h$ and $ {\cal H}$ belong to the redundant field if one chooses
the perturbations along Eqs.(\ref{ptr_metric})-(\ref{ptr_scalar}).
It confirms that our counting for degrees of freedom is correct.
We note that the harmonic gauge with the Kalb-Ramond equation (\ref{lin_anti})
leads to the same constraint as in Eq.(\ref{eq_anti_simple}).
As a result, Eq.(\ref{eq_scalar}) becomes a decoupled dilaton equation
\begin{equation}
\bar \nabla^2 \varphi + \fr{8}{\ell^2} \varphi=0
\label{deq}
\end{equation}
which can be rewritten explicitly as

\begin{eqnarray}
\left [ f^2 \partial_r^2
+ \left\{ {1 \over r} (\partial_r rf^2) \right\} \partial_r
-{{J j \omega} \over {r^2 f^2}}
+{\omega^2 \over f^2}
+{(-M+{r^2 /\ell^2)} \over r^2 f^2} j^2
\right ] \tilde \varphi
+{8 \over l^2}\tilde \varphi
=0,
\label{eq_decoupled}
\end{eqnarray}
It is noted that if the last term is absent, Eq.(\ref{eq_decoupled})
corresponds to the wave equation of a free scalar in Kerr-de Sitter space.
Comparing this dilaton equation with the massive scalar
equation
\begin{equation}
\bar \nabla^2 \phi_m -m^2 \phi_m=0,
\label{meq}
\end{equation}
it seems  that the dilaton propagates on  Kerr-de Sitter space
with the tachyonic mass $m^2_{\varphi}=-8/\ell^2$.
However, this observation is not the whole of story in de Sitter space.

\section {stability analysis for de Sitter space}

It is not easy to solve the wave equation  Eq.(\ref{eq_decoupled}) of the dilaton
on the southern diamond  including  the cosmological horizon ($r=r_c$) and the origin ($r=0$).
 The main difficulty comes from the fact that the
black hole horizon degenerates to give a conical singularity at
$r=0$. In other words, the Kerr-de Sitter solution represents a spinning point mass
$M$ in de Sitter space. This  makes it hard to express the solution to
the wave equation on the southern diamond in terms of the
hypergeometric function. Then we cannot calculate the dilaton  absorption
cross section to test  the cosmological horizon. Hence, hereafter, we
confine ourselves to the pure  de Sitter solution with $J=0$. Then the
origin is just that of the coordinate and thus there is nothing to worry about
singularity on the southern diamond. Even though we give up
the Kerr-de Sitter background, we can study the nature of de
Sitter space using the dilaton.

Eq.(\ref{eq_decoupled}) reduces to the differential equation for
$r$\cite{BMS}
\beq
(1-r^2)\tilde \varphi''(r) +( \fr{1}
{r} -3r) \tilde \varphi'(r) + \Big( \fr{\omega^2}{1-r^2}
-\fr{j^2}{r^2} -m_\varphi^2 \Big) \tilde
\varphi(r)=0,~~m_\varphi^2=-8
\label{2eq4}
\eeq
where the prime ($'$) denotes the differentiation with respect to
its argument and for simplicity we take $\ell$ to be 1. The
original equation from (\ref{eq_decoupled}) with $J=0, \ell \not=1$
takes the same form as in Eq.(\ref{2eq4}) if
$r/\ell\to \tilde r, \omega \ell \to \omega, m_\varphi \ell \to
m_\varphi$. This information will be used for obtaining the
absorption cross section in section V.
From Eq.(\ref{2eq4}) it is obscure to know how the dilaton wave propagates in the southern
diamond. In order to show it clearly, we must transform the wave equation
into the Schr\"odinger-like equation by introducing a tortoise
coordinate $r^*$.  Then we can get
information through the potential analysis.
We introduce  $r^*=g(r)$ with $ g'(r)=1/r(1-r^2)$ to transform  Eq.(\ref{2eq4})
into the Schr\"odinger-like equation with the energy
$E=\omega^2$\cite{YS}

\beq
-\fr{d^2}{d r^{*2}}\varphi + V_{ \varphi}(r) \tilde \varphi=  E \tilde \varphi
\label{3eq1}
\eeq
with the potential
\beq
V_{\varphi}(r)= \omega^2 + r^2(1-r^2) \Big[ m_{\varphi}^2 + \fr{j^2}{r^2} -\fr{\omega^2}{1-r^2}
\Big].
\label{3eq2}
\eeq
Considering $r^*=g(r)= \int g'(r) dr$, one finds
\beq
r^*= \ln r - \fr{1}{2} \ln(1-r^2),~~ e^{2r^*} =\fr{r^2}{1-r^2},~~
r^2=\fr{e^{2r^*}}{1+e^{2r^*}}.
\label{3eq3}
\eeq
Here we confirm that $r^*$ is a tortoise
coordinate such that $r^* \to -\infty (r \to 0)$, whereas $r^* \to \infty (r \to
1)$. Let us express the potential as a function of $r^*$
\beq
V_{ \varphi}(r^*) = \omega^2 + \fr{e^{2r^*}}{(1+e^{2r^*})^2} \Big[ m_{ \varphi}^2
 +
\fr{1+ e^{2r^*}}{e^{2 r^*}} j^2 - (1 + e^{2r^*}) \omega^2 \Big].
\label{3eq4}
\eeq
First of all we mention that this potential is the
energy-dependent potential. Let us consider the low-energy limit
of $\omega^2 \ll 1$.
For $m_{\varphi}^2=-8,j=0,\omega=0.1$, the shape of this takes a  potential well
near $r^*=0$. Due to the potential well, this potential induces an
exponentially large dilaton  which is obviously  contradicted to the
genuine small value of the perturbation. Hence the $j=0(s)$-mode of the dilaton
seems to be
unstable. Naively speaking, this means that the cosmological horizon in our model
does not truly exist. However we have some ambiguity to define this $s$-mode in de Sitter space.
Hence it is not clear from $s$-mode analysis that the cosmological horizon is unstable.

 For $j\not=0$-modes with the low-energy of $\omega^2 \ll 1$, one finds that
 $V_{\varphi}(r^*=0)= -8 + j^2/2 +\omega^2/2 $. Hence for
 $j=1,2,3$-modes one finds the potential wells,  which confirm that
 the cosmological horizon is unstable. For $j\ge 4$-modes the
 potential well disappears.
In addition  one finds the
potential step with its height $\omega^2+j^2$ on the left-hand side.
 All potentials decrease exponentially to zero as $r^*$
increases on the right-hand side.
This means that we always develop  a well-defined wave near
the cosmological horizon of $r^*=\infty$. But near $r^*=-\infty (r=0)$
it is not easy to develop the genuine waves. It is expected that
the scattering to give a finite absorption cross section will occur if
 $E=\omega^2 \sim V_{ \varphi}(r^*)$.
 This case is possible if  $\omega^2 \gg j^2$.
 This means that the relevant scattering in de Sitter space may be
 arisen if the frequency of the external field is larger than its
 angular momentum quantum number. This corresponds to the
 low-temperature limit of $\omega > T_c$. In this case the
 scattering can be defined even for $j=1,2,3$ cases.

\section{absorption cross section}

The absorption coefficient by the cosmological horizon is defined by
 the ratio of the outgoing flux at $r=0$ to the outgoing flux at $r=r_c$
 as
\beq
{\cal A} = \fr{ {\cal F}_{out}(r=r_c)}{{\cal F}_{out}(r=0)}.
\label{5eq1}
\eeq
Up to now we do not
insert the curvature radius $\ell$ of dS$_3$ space.
The correct absorption coefficient can be recovered by replacing $\omega(m_\varphi)$ with
$ \omega\ell(m_\varphi \ell)$. Here we do not repeat the procedure of the flux
computation but refer ref.\cite{YS}.
Then the dilaton absorption cross section in three dimensions is defined
by
\beq
\sigma_{abs}= \fr{{\cal A}}{\omega} =
\Big[\fr{A_r^2+B_i^2}{A_rB_i}\Big] \fr{\ell}{j}|\alpha_{-\omega\ell,j}|^2
\label{5eq2}
\eeq
where
\beq
|\alpha_{-\omega\ell,j}|^2=\fr{|\Gamma(1+j)|^2|\Gamma(i\omega\ell)|^2 }
{|\Gamma[2+j/2 + i\omega\ell/2)]|^2
|\Gamma[j/2 + i\omega\ell/2]|^2}.
\label{5eq3}
\eeq
We observe that $ s(j=0)$-mode cross section is
ill-defined because ${\cal F}_{out}(r=0)=0$ for $j=0$.
Furthermore the normalization factor of $ \Big[\fr{A_r^2+B_i^2}{A_rB_i}\Big] $
is not fixed by the theory. In order to obtain the explicit form, let us calculate
$|\alpha_{-\omega\ell,j}|^2$ according to values of the angular momentum quantum number $j$.
One finds
\beqa
\sigma_{abs,j}^{dil}&=&\Big[\fr{A_r^2+B_i^2}{A_rB_i}\Big] \fr{\ell}{j}
\fr{(j!)^2\pi}{ \omega\ell \sinh[\pi\omega\ell]}\fr{16}{((2+j)^2+(\omega\ell)^2)
 (j^2+(\omega\ell)^2)}\nonumber \\
 && \times \fr{1}{|\Gamma(j/2 + i\omega\ell/2)|^2|\Gamma(j/2 + i\omega\ell/2)|^2}.
\label{5eq4}
\eeqa
For $j=1$, this takes the form
\beqa
\sigma_{abs,j=1}^{dil}&=&\Big[\fr{A_r^2+B_i^2}{A_rB_i}\Big]
\fr{8 {\cal A}_{ch}}{ \pi^2 \omega\ell \sinh[\pi\omega\ell]} \nonumber \\
 && \times \fr{\Big(\cosh[\pi\omega\ell/2]\Big)^2}{(9+ (\omega\ell)^2)(1+(\omega\ell)^2)}
\label{5eq5}
\eeqa
with the area of the cosmological horizon ${\cal A}_{ch}=2 \pi
\ell$.
In order to get a definite expression for the absorption cross section,
let us consider the low-energy scattering with small $E=(\omega\ell)^2 \ll1$
(for example, $\omega\ell=0.1$). Noting the temperature of
the cosmological horizon $T_c=\fr{1}{2\pi\ell}$, this limit implies
$\omega<T_c$. On the other hand, there exists the low-temperature
limit of $(\omega\ell)^2 \gg 1(\omega>T_c)$.
In the low-energy limit of $\omega\ell < 1$, $j=1$-mode reduces to
\beq
\sigma_{abs,j=1}^{dil,\omega\ell<1}=\Big[\fr{A_r^2+B_i^2}{A_rB_i}\Big]
\fr{8}{ 9\pi^3} \fr{{\cal A}_{ch}}{(\omega\ell)^2}.
\label{5eq6}
\eeq
On the other hand, its low-temperature limit  is given by
\beq
\sigma_{abs,j=1}^{dil,\omega\ell>1}=\Big[\fr{A_r^2+B_i^2}{A_rB_i}\Big]
\fr{8}{ \pi^2} \fr{{\cal A}_{ch}}{(\omega\ell)^5}.
\label{5eq6'}
\eeq

For $j=2$, this leads to
\beqa
\sigma_{abs,j=2}^{dil}&=&\Big[\fr{A_r^2+B_i^2}{A_rB_i}\Big]
\fr{16 {\cal A}_{ch}}{\omega\ell \sinh[\pi\omega\ell]} \nonumber \\
 && \times \fr{\Big(\sinh[\pi\omega\ell/2]\Big)^2}
 {(16+(\omega\ell)^2)(4+(\omega\ell)^2)(\pi\omega\ell/2)^2}.
\label{5eq7}
\eeqa
In the low-energy limit of $\omega\ell < 1$, this reduces to
\beq
\sigma_{abs,j=2}^{dil,\omega\ell<1}=\Big[\fr{A_r^2+B_i^2}{A_rB_i}\Big]
\fr{1}{16 \pi} \fr{{\cal A}_{ch}}{(\omega\ell)^2}.
\label{5eq8}
\eeq
But its low-temperature limit is given by
\beq
\sigma_{abs,j=2}^{dil,\omega\ell> 1}=\Big[\fr{A_r^2+B_i^2}{A_rB_i}\Big]
\fr{64}{ \pi^2} \fr{{\cal A}_{ch}}{(\omega\ell)^7}.
\label{5eq8'}
\eeq

\section{discussion}
We calculate the absorption cross section of the dilaton
 which propagates on the southern diamond of
three-dimensional de Sitter space.
One of the striking results is that the low-energy $s(j=0)$-wave
absorption of the dilaton is not defined properly.  This
mainly rests on being  unable to calculate its finite flux at
$r=0$. This contrasts sharply to the cases found in the symmetric black holes
whose $s$-wave cross sections are well-defined and
proportional to the area of the event horizon\cite{grey2}.

On the other hand, the $j\not=0$-angular modes of the dilaton
can be used for exploring  the dynamical aspects of the
cosmological horizon.
We expect  from the black hole analysis that
the low-energy limit ($\omega R \to 0 $) of the $l\not=0$-angular mode absorption cross section
is proportional roughly to $(\omega R)^{4l}$ for the  D=7 black hole which
is induced from D3-branes\cite{GH}. For D=5 black hole, it is
proportional to $(\omega r_o)^{2l}$\cite{MS}.
 However, one finds from Eqs.(\ref{5eq6}) and (\ref{5eq8})
 that those for $j\not=0$ in the low-energy limit of $\omega\ell<1$ are given by
 $(\omega\ell)^{-2}$
which implies that
the absorption cross section is greater than the area of the cosmological horizon.
This is not the case what we want to get.
From the potential analysis in section IV,
it conjectures that for $j=1,2,3$ cases,  the low-energy absorption cross section
with $E=(\omega\ell)^2 \ll 1$
is meaningless  because these become the unstable case.
This implies that to obtain the
finite absorption cross section,
$\omega\ell$ should be large such as $\omega\ell > j$. This corresponds to
  the low-temperature limit of $\omega> T_c$.
The low-temperature limit
 is  meaningful in de Sitter space since its cross section appears less than ${\cal A}_{ch}$.
For example, we find from Eqs.(\ref{5eq6'}) and  (\ref{5eq8'})
that the absorption cross sections takes
$\sigma_{abs,j}^{dil,\omega\ell>1} \sim (\omega\ell)^{-(2j+3)}$ roughly.
This is consistent with the potential analysis. According to the
this, the potential height is proportional to $\omega^2+j^2$ which
implies that  the absorption cross section decreases as $j$ increases.
On the other hand, the free scalar absorption cross section takes
the same form $(\omega\ell)^{-2}$ as in the dilaton in the low-energy limit, while it
is
given by $\sigma_{abs,j}^{free,\omega\ell>1} \sim (\omega\ell)^{-(2j+1)}$
 in the low-temperature limit\cite{YS}.
As a result, we confirm that  the low-temperature limit (not the low-energy
limit) of
$j\not=0$-angular mode absorption cross section will be used to
test the cosmological horizon in de Sitter space.
This contrasts sharply to the fact that the low-energy $s$-mode
plays an important role to test the black hole event horizon.

In conclusion, to get information about the cosmological horizon,
 we have to inject the test field with high
frequency into the de Sitter background.

Finally we mention that the AdS bulk absorption cross section can be
also calculated from the two-point function of CFT defined on the boundary if one assumes the
AdS/CFT correspondence using the boundary-bulk Green function\cite{TEO}.
Hence we propose that our results for the dS bulk space can be recovered from
the Euclidean CFT by
making use of the dS/CFT correspondence \cite{STR,SV} and the corresponding
boundary-bulk Green function\cite{AWLQ}.

\section*{Acknowledgement}

This work was supported in part by the Brain Korea 21
Program of  Ministry of Education, Project No. D-1123.

\end{document}